\documentclass[twocolumn]{article}

\usepackage{PRIMEarxiv}
\usepackage{algorithm}
\usepackage[utf8]{inputenc} % allow utf-8 input
\usepackage[T1]{fontenc}    % use 8-bit T1 fonts
\usepackage{hyperref}       % hyperlinks
\usepackage{url}            % simple URL typesetting
\usepackage{booktabs}       % professional-quality tables
\usepackage{amsmath,amssymb,amsfonts}       % blackboard math symbols
\usepackage{nicefrac}       % compact symbols for 1/2, etc.
\usepackage{microtype}      % microtypography
\usepackage{lipsum}
\usepackage{fancyhdr}       % header
\usepackage{graphicx}       % graphics
\graphicspath{{media/}}     % organize your images and other figures under media/ folder
\usepackage{cite}
\usepackage{algorithmic}
\usepackage{graphicx}
\usepackage{textcomp}
\newtheorem{theorem}{Theorem}

\title{Efficient Pairing in Unknown Environments: Minimal Observations and TSP-based Optimization\footnote{This work was supported in part by the Japan Science and Technology Agency through the Core Research for Evolutionary Science and Technology (CREST) Project under Grant JPMJCR17N2, and in part by the Japan Society for the Promotion of Science through the Grants-in-Aid for Scientific Research (A) under Grant JP20H00233.}}
\date{}
\author{
Naoki Fujita\footnote{Department of Information Physics and Computing, Graduate School of Information Science and Technology, The University of Tokyo, Hongo, Bunkyo-ku, Tokyo 113-8656, Japan}\and
Nicolas Chauvet\footnotemark[2]\and
Andr\'e R\"{o}hm\footnotemark[2] \and
Ryoichi Horisaki\footnotemark[2]\and
Aohan Li\footnote{Department of Electrical Engineering, Graduate School of Engineering, Tokyo University of Science, 6-3-1 Niijuku, Katsushika-ku, Tokyo 125-8585, Japan}\and
Mikio Hasegawa\footnotemark[3]\and
Makoto Naruse\footnotemark[2]}

\begin{document}
\maketitle
\begin{abstract}
Generating paired sequences with maximal compatibility from a given set is one of the most important challenges in various applications, including information and communication technologies. However, the number of possible pairings explodes in a double factorial order as a function of the number of entities, manifesting the difficulties of finding the optimal pairing that maximizes the overall reward. In the meantime, in real-world systems, such as user pairing in non-orthogonal multiple access (NOMA), pairing often needs to be conducted at high speed in dynamically changing environments; hence, efficient recognition of the environment and finding high reward pairings are highly demanded. In this paper, we demonstrate an efficient pairing algorithm to recognize compatibilities among elements as well as to find a pairing that yields a high total compatibility. The proposed pairing strategy consists of two phases. The first is the observation phase, where compatibility information among elements is obtained by only observing the sum of rewards. We show an efficient strategy that allows obtaining all compatibility information with minimal observations. The minimum number of observations under these conditions is also discussed, along with its mathematical proof. The second is the combination phase, by which a pairing with a large total reward is determined heuristically. We transform the pairing problem into a traveling salesman problem (TSP) in a three-layer graph structure, which we call Pairing-TSP. We demonstrate heuristic algorithms in solving the Pairing-TSP efficiently. This research is expected to be utilized in real-world applications such as NOMA, social networks, among others.
{\flushleft{{\bf Keywords:} Pairing; exchange rule; traveling salesman problem; maximum weighted matching; combinatorial optimization; compatibility; non-orthogonal multiple access}}
\newline
\end{abstract}

%Bibliography
\bibliographystyle{unsrt}  
\bibliography{references}  

\section{Introduction}
\subsection{Introduction of a pairing problem}
Various systems and applications require to combine multiple elements into an array of pairs, including information or communication technologies. The process of partitioning the set of $N$ elements into $N/2$ disjoint sets with exactly 2 elements each is called "Pairing" in this paper. An example is found in non-orthogonal multiple access (NOMA) in the latest wireless communication systems \cite{NOMA,NOMA2,NOMA3,NOMA4,NOMA5,NOMA6}. In NOMA, multiple terminals share a common frequency band simultaneously, which greatly improves the frequency utilization efficiency. The key process here is user pairing: the base station allocates higher and lower transmission power for communications to the terminals located far and near the base station, respectively. The terminals then conduct 
successive interference cancellation (SIC) calculations to extract the original signal.
Therefore, determining the combination of user pairing that maximizes the total data rate of all users is critical. However, to the best of the authors’ knowledge, optimal pairing algorithms which can work with a large number of users or terminals have not been proposed, even though various pairing algorithms have been proposed in previous studies\cite{NOMA7,NOMA8}. When the number of users $N=10$, the total number of possible pairings is 945. With $N=100$, the total number becomes in the order of $10^{78}$, which is a double factorial scaling as introduced in Sect. III. Therefore, an efficient pairing strategy is indispensable.
The importance of pairing is also observed in other situations and applications, such as college admission \cite{college}, economics \cite{economics} and donor exchange\cite{donor} among others\cite{air,team}. 

In this paper, we demonstrate a fast pairing algorithm consisting of an efficient recognition of compatibilities among elements as well as an efficient determination of the pairing that yields high total compatibility. Here, compatibility quantifies the performance of a given pair and total compatibility is the summation of compatibilities of all pairs for a pairing, where we also call the given set of pairs among all elements a pairing. The optimal pairing should maximize the total compatibility of the system. 
However, in general, obtaining the globally maximal total compatibility would require an exhaustive search of all pairings. Therefore, a heuristic algorithm is needed to obtain an approximately maximal total compatibility. This study highlights the following two aspects in discussing the pairing problem. 

\begin{figure}[h]
\centering
\includegraphics[width=8cm]{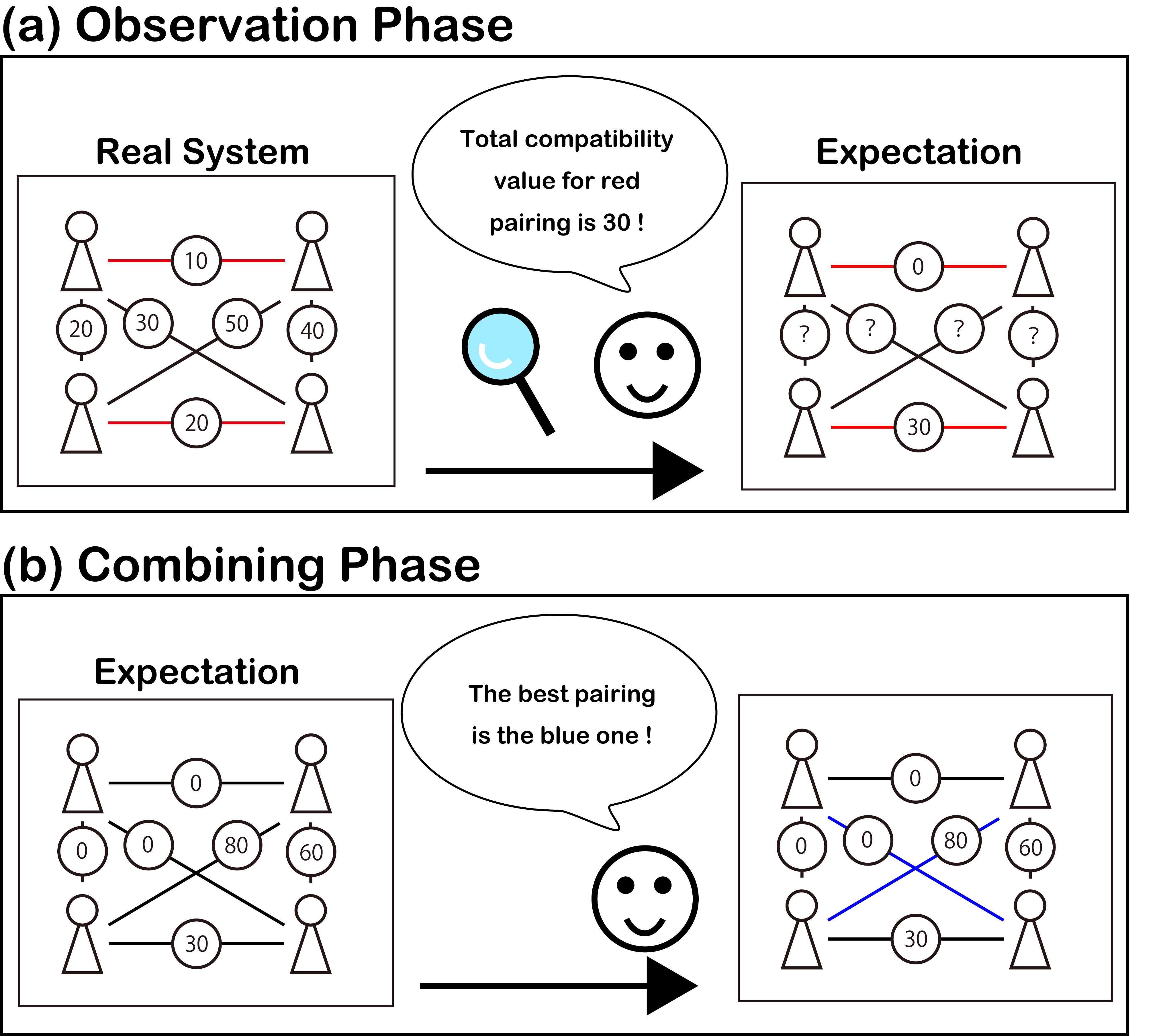}
\caption{Efficient pairing in unknown environments. There are two phases. (a) The first is the \textit{observation phase} to grasp the compatibility among elements. (b) The second is the \textit{combining phase} to find a pairing yielding high compatibility.}
\label{overview}
\end{figure}

The first point is the time duration required to obtain information about the compatibilities of the system, which we call \textit{observation time} hereafter. In the absence of prior information about compatibilities, multiple observations are required to infer the compatibility between all elements. 
Furthermore, we presuppose that we cannot directly measure individual compatibility among elements; only the total compatibility of a certain pairing is observable.
The fewer observations, the shorter the overall time required for pairing. More generally, the objective of an algorithm for compatibility observation is to guess as accurately as possible the real compatibilities in as few steps as possible, which is schematically illustrated in Fig. \ref{overview}(a). In this paper, we demonstrate that by exploiting the inherent structural properties of the pairing problem, which we call exchange rules, the number of observations needed for acquiring all compatibilities is significantly reduced. 

The second point is the efficient derivation of the optimal pairing based on the information on compatibility; we call the required time for this process \textit{combining time} in this paper. 
Even if complete information about the compatibilities is available, it may take a considerable amount of time to find the optimal pairing because of the huge number of possible pairings, as schematically represented in Fig. \ref{overview}(b). In this paper, we transform the derivation of optimal pairing into a traveling salesman problem (TSP). TSP is a widely known combinatorial optimization problem to find the shortest pathway in a graph $G(V, E)$ for a salesman while visiting all vertices $V$ via edges $E$. In addition to the compatibility information, we append two more layers to account for the requirement of the pairing problem; we call the re-formulated pairing problem the Pairing-TSP. Notably, the resulting graph is not fully connected. Once the situation is represented by a TSP problem, we can benefit from a variety of heuristic algorithms in the literature to efficiently solve the combinatorial explosion issue. 
Furthermore, this paper proposes a novel heuristic algorithm that is different from conventional algorithms and suitable for pairing problems.

Regarding the second point discussed above, a related problem is maximum weight matching (MWM). In MWM, the goal is to select edges from a weighted graph $G(V, E)$ so that any two selected edges do not share common vertices while maximizing the sum of the weights of the selected edges. This is a combinatorial optimization problem. The pairing problem discussed in this study is a particular case of MWM where the graph is complete and the number of vertices is even. Several efficient algorithms have been proposed for solving MWM. For example, Gabow\cite{MWM} proposed an algorithm with a computation time of $|E||V|+|V|^2\log{|V|}$, Cygan \textit{et al}.\cite{RMWM} developed a randomized algorithms with a computation time of $L|V|^\omega$ for graphs with integer weights ($\omega<2.373$ is the exponent of $N \times N$ matrix multiplication\cite{omega} and $L$ is the maximum integer edge weight), while  Duan \textit{et al}. \cite{AMWM} worked on an algorithm achieving an approximation ratio of $(1-\epsilon)M$, with computation time of $|E|\epsilon^{-1}\log{\epsilon^{-1}}$ for arbitrary weights and $|E|\epsilon^{-1}\log{N}$ for integer weights ($\epsilon$ is a positive arbitrary value and $M$ is the maximum value). Here, $|V|=N, |E|=N(N-1)/2$.

In our paper, we approach the problem from a different perspective and present new methods for the pairing problem. In particular, we formulate it as a TSP problem. Additionally, MWM literature does not consider any approach to obtain compatibility information to the best of our knowledge, which is the main aspect of the observation phase in our manuscript.

\subsection{Overview of this paper}
With a view to the efficient realization of optimal pairing, the present study demonstrates an efficient observation strategy to measure the compatibilities among the entities on the basis of limited information. Furthermore, based on the insight that the optimal pairing problem is transformed into a TSP problem in a three-layer graph structure, we demonstrate heuristic algorithms to find a high-performance pairing that can be applied even when the number of elements is large.

This paper is organized as follows. First, we formulate the pairing problem in Sect. II. Second, for the observation phase, in Sect. III, we show the minimum number of observations needed to infer the complete set of compatibilities and propose an observation algorithm with a computational complexity of the square of the number of elements. Sect. IV examines the combining phase, where we introduce how to convert the pairing problem to a TSP and propose an algorithm for solving the resulting Pairing-TSP. In Sect. V, we numerically evaluate the performances of the combining phase algorithms. Sect. VI concludes the paper.

\section{Objective function and constraints}
Here we assume that the number of elements is an even natural integer $N$, while the index of each element is a natural number between 1 and $N$. We define the set of all users $\mathbb{U}$ as follows;
\begin{equation*}
    \mathbb{U}=\{i | 1\leq i \leq N\}.
\end{equation*}
Then, we define the set of all possible pairs for $\mathbb{U}$ as $\mathbb{D}$:
\begin{equation*}
    \mathbb{D}=\{\{i,j\}| 1\leq i < j \leq N\}\\
\end{equation*}
The compatibility between the elements $i$ and $j$ is denoted by $C_{i,j}$. The reward function $f$ of a pair is given by its compatibility. 
\begin{equation*}
    f(\{i,j\})=C_{i,j}.\\
\end{equation*}
We define pairing $S$ as follows;
\begin{eqnarray*}
    &&\forall d \in S, d\in \mathbb{D}\\
    &&\bigcup S = \mathbb{U}\\
    &&A,B\in S, A\neq B\Rightarrow A\cap B = \emptyset
\end{eqnarray*}
Then, $C_{\rm{sum}}({S})$, which is called "the total compatibility of pairing $S$" hereafter is defined as follows;
\begin{equation*}
    C_{\rm{sum}}(S)=\sum_{d\in S}f(d).
\end{equation*}
And we define the set of all pairings $\mathbb{S}=\{S\}$. The pairing problem discussed in this study is formulated as follows;
\begin{eqnarray*}
    &&\textrm{max:}\,C_{\rm{sum}}(S)\\
    &&\textrm{subject\,to:}\,S \in \mathbb{S}.
\end{eqnarray*}

\section{Observation Phase}
\subsection{Exchange Rule}
 As discussed in the Introduction, we assume that each $C_{i,j}$ cannot be directly observed, but $C_{\rm{sum}}(S)$ is observable. By observing such $C_{\rm{sum}}(S)$ values with different pairings $S$, we can recognize all $C_{i,j}$ values.
The number of all available pairings is $(N-1)!!$, meaning that the number of necessary observations is at most $(N-1)!!$. Here, the number denoted by $n!!$ is the double factorial of an odd number $n$ defined by $n(n-2)(n-4) \cdots 3 \cdot 1$. Therefore, the total number of possible pairings dramatically increases when $N$ becomes large, indicating the importance of efficiently recognizing compatibilities with as few observations as possible. 
In the following, we prove that the total compatibility $C_{\rm{sum}}(S)$ of all possible pairings can be calculated based on a limited number of observations, leading to a significant reduction of the required observations. 

To improve the readability of the following discussion, we define the {\it exchange rule} as:
\begin{equation}
[i,j,k,l]\equiv (C_{i,k}+C_{j,l})-(C_{i,j}+C_{k,l}).
\end{equation}
This exchange rule describes the amount of change in the total compatibility between a pairing $S$ containing $\{i, j\}$ and $\{k, l\}$ and a pairing $S$ containing $\{i, k\}$ and $\{j, l\}$.
Therefore, each exchange rule can be calculated from two observations. For example, to find $[1,2,3,4]$ in $N=8$, we can observe $C_{1,2}+C_{3,4}+C_{5,6}+C_{7,8}$ and $C_{1,3}+C_{2,4}+C_{5,6}+C_{7,8}$ and calculate the difference.
For a large $N$, there are many sets of pairings corresponding to any given exchange rule, such that finding one exchange rule will give the amount of change between multiple sets simultaneously. For example, $[1,2,3,4]$ mentioned above is also the difference between $C_{1,2}+C_{3,4}+C_{5,7}+C_{6,8}$ and $C_{1,3}+C_{2,4}+C_{5,7}+C_{6,8}$.

\subsection{Observation Algorithm}
In this part we propose a simple algorithm with observation time in $\mathcal{O}(N^2)$. As an example, we will use the $C_{i,j}$ setting shown in Table \ref{first_setting} to illustrate the proposed observation algorithm.
As discussed earlier, we assume that each compatibility ($C_{i,j}$) cannot be observed directly. It will prove beneficial to not calculate the original set of compatabilities $C_{i,j}$ directly, but to use a derived set of compatiabilities denoted by $\tilde{C}_{i,j}$ and $\tilde{C}_{\rm{sum}}(S)$ with the two following properties: for a given pairing, $C_{\rm{sum}}(S)=\tilde{C}_{\rm{sum}}(S)$, and some $\tilde{C}_{i,j}$ are always equal to 0. If such properties hold, we could calculate any total compatibility via $\tilde{C}_{i,j}$ with a reduced number of observations, instead of via $C_{i,j}$.

Indeed, we found that such $\tilde{C}_{i,j}$ exists and can be defined as the following;
\begin{equation}
\tilde{C}_{i,j}=
\begin{cases}
0,\, \text{if $i=1$ or $j=1$,}\\
C_{i,j}-C_{1,i}-C_{1,j}+\frac{2}{N-2}\sum_{k=2}^{N}C_{1,k},\,\text{otherwise.}\nonumber
\end{cases}
\end{equation}
In this definition, we found that the number of non-zero $\tilde{C}_{i,j}(i>j)$ elements is $(N-1)(N-2)/2$, which is smaller than the number of $C_{i,j}(i>j)$ elements. That is, this definition reduces the number of non-zero elements.
When we denote a pairing as $S$, from the definition of $\tilde{C}_{i,j}$ we can write;
\begin{eqnarray*}
&&\tilde{C}_{\rm{sum}}(S)\\
&=&\sum_{\{i,j\}\in S}\tilde{C}_{i,j}\\
&=&\tilde{C}_{1,l}+\sum_{\{i,j\}\in S\setminus \{1,l\}}\tilde{C}_{i,j}\\
&=&\sum_{\{i,j\}\in S\setminus \{1,l\}}\tilde{C}_{i,j}\\
&=&\sum_{\{i,j\}\in S\setminus \{1,l\}}\left(C_{i,j}-C_{1,i}-C_{1,j}+\frac{2}{N-2}\sum_{k=2}^{N}C_{1,k}\right)\\
&=&\sum_{\{i,j\}\in S\setminus \{1,l\}}C_{i,j}-\sum_{2\leq k \leq N, k\neq l} C_{1,k}+\sum_{k=2}^{N}C_{1,k}\\
&=&\sum_{\{i,j\}\in S\setminus \{1,l\}}C_{i,j}+C_{1,l}\\
&=&\sum_{\{i,j\}\in S}C_{i,j}\\
&=&C_{\rm{sum}}(S).
\end{eqnarray*}
The above equations prove that $C_{\rm{sum}}(S)$ and $\tilde{C}_{\rm{sum}}(S)$ are equal for any pairing $S$.\\
\indent As a consequence, any exchange rule can be written using either $C_{i,j}$ or $\tilde{C}_{i,j}$ while providing the same value. Thanks to this property and $\tilde{C}_{1,j}=0$ for any $j$, the computation can be greatly simplified. For example, if we compute the value of the exchange rule $[1,i,j,k]$ for $\tilde{C}_{i,j}$, we can transform the equation as follows;
\begin{eqnarray}
[1,i,j,k]&=&(\tilde{C}_{1,j}+\tilde{C}_{i,k})-(\tilde{C}_{1,i}+\tilde{C}_{j,k})\nonumber\\
&=&\tilde{C}_{i,k}-\tilde{C}_{j,k}.\nonumber
\end{eqnarray}
That is, we can obtain the difference between the two elements ($\tilde{C}_{i,k}$ and $\tilde{C}_{j,k}$) from a single exchange rule.\\
\indent In the proposed observation algorithm, the following values (Eqs. (2), (3), and (4)) are obtained from observations;
\begin{equation}
[1,j,3,2]\,(4\leq j \leq N),
\end{equation}
\begin{equation}
[1,i,2,j]\,(4\leq j \leq N, 3\leq i\leq j-1),
\end{equation}
\begin{equation}
\sum_{i=1}^{N/2}C_{2i-1,2i}.
\end{equation}
By definition, following equations hold;
\begin{eqnarray}
[1,j,3,2]&=&(\tilde{C}_{1,3}+\tilde{C}_{2,j})-(\tilde{C}_{1,j}+\tilde{C}_{2,3}) \nonumber \\
&=&\tilde{C}_{2,j}-\tilde{C}_{2,3},
\end{eqnarray}
\begin{eqnarray}
[1,i,2,j]&=&(\tilde{C}_{1,2}+\tilde{C}_{i,j})-(\tilde{C}_{1,i}+\tilde{C}_{2,j}) \nonumber \\
&=&\tilde{C}_{i,j}-\tilde{C}_{2,j}.
\end{eqnarray}
\begin{table}[h]
\centering
\caption{The original $C_{i,j}$ setting}
\includegraphics[width=5cm]{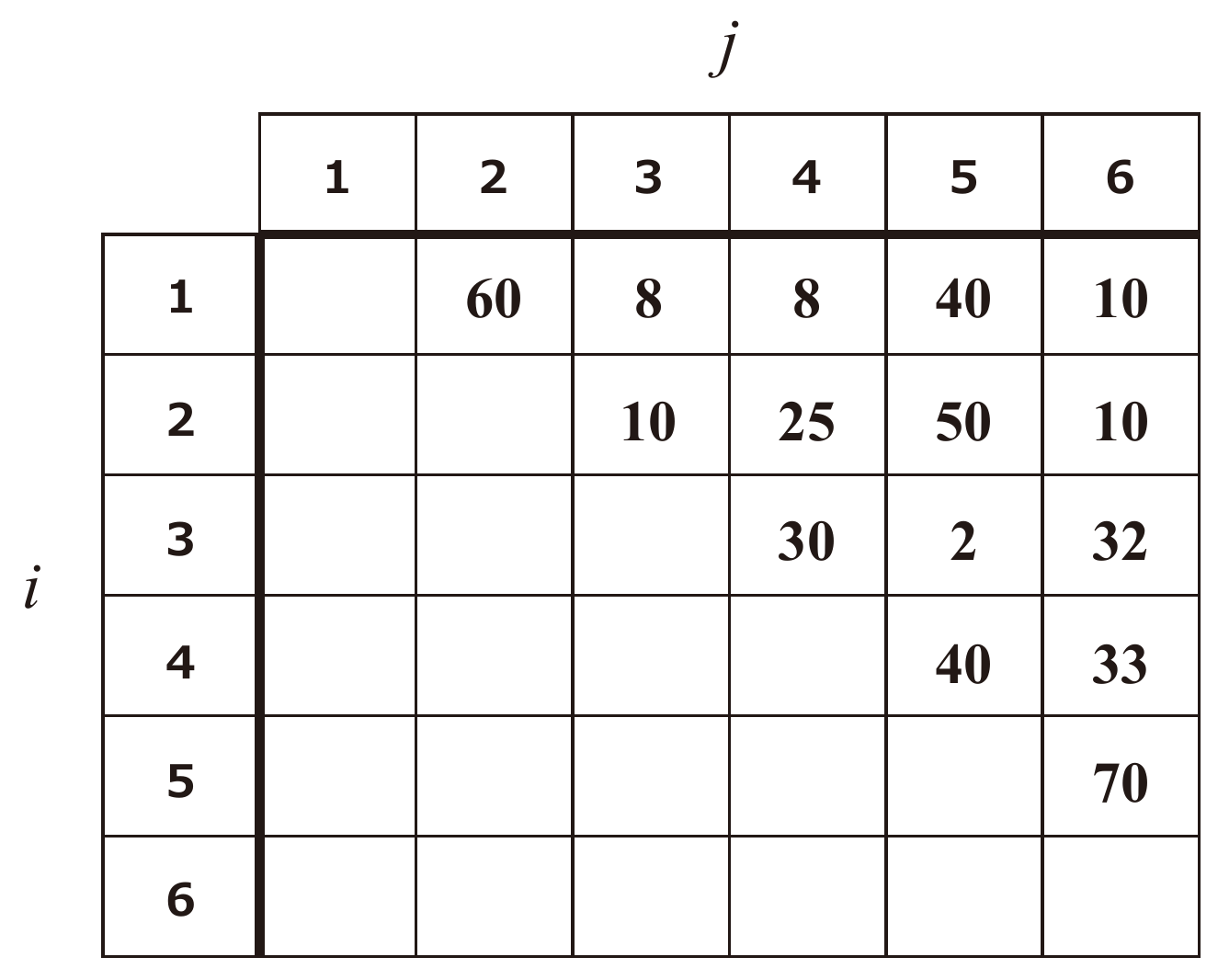}
\label{first_setting}
\end{table}
\begin{table}[h]
\caption{(a) The change in the horizontal direction $i=2$ (solid arrow), and the change in the vertical direction $i$ (dotted arrow). (b) $\tilde{C}_{i,j}$ calculated by observation.}
\centering
\includegraphics[width=8cm]{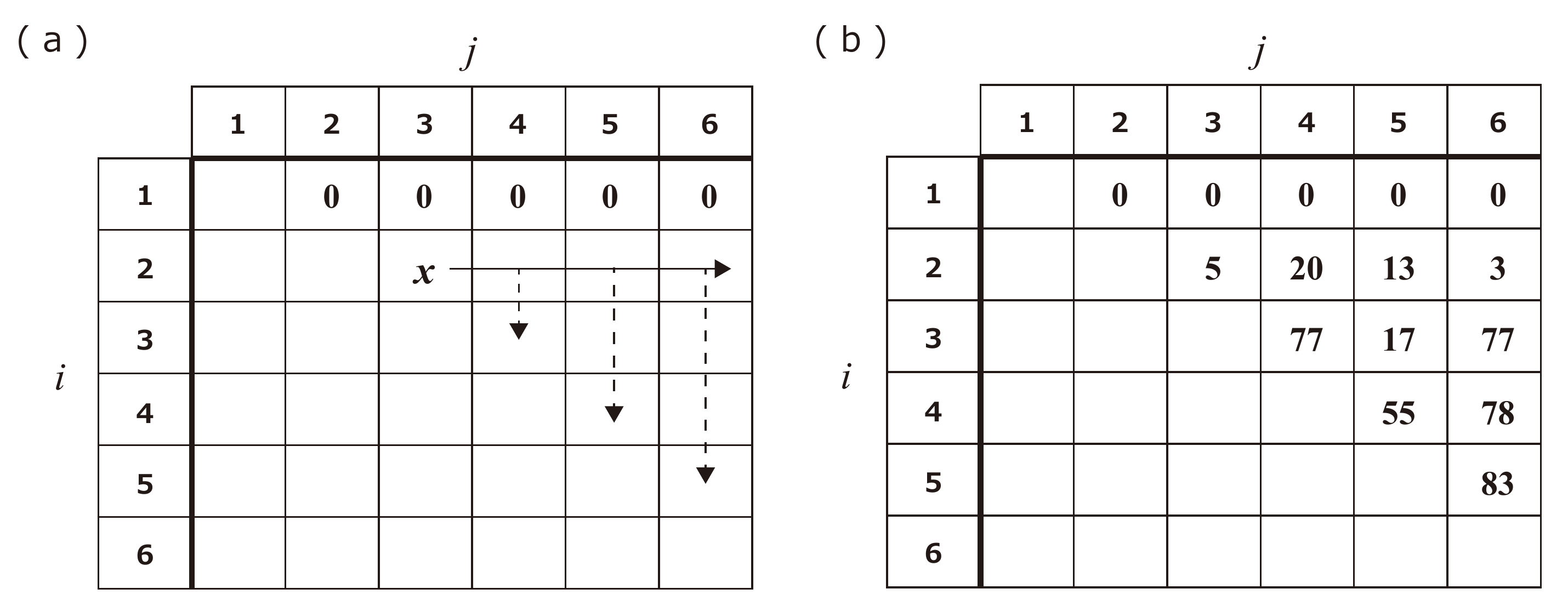}
\label{second_setting}
\end{table}
Eq. (5) represents the changes along the horizontal direction $i = 2$, and Eq. (6) represents the changes in the vertical directions $i$ (in Table \ref{second_setting}(a)). 
Let $C_{2,3}$ be given by $x$, $\tilde{C}_{i,j}$ is represented using Eqs. (5), and (6). 
Then, by using Eq. (4), $x$ is determined, and subsequently all $\tilde{C}_{i,j}$ values are determined as summarized in Table \ref{second_setting}(b). 
Number of observations needed for Eqs. (2), (3), (4) are $2(N-3), (N-2)(N-3)$, and 1, respectively, because each exchange rule value can be calculated by only two observations. It follows that the observation number by the proposed algorithm is $\mathcal{O}(N^2)$.

\subsection{Minimum Number of Observations}
We proved the following theorem;
\\
\begin{theorem}
The minimum number of observations required to know the entire set of compatibilities $\tilde{C}$ is $(N-1)(N-2)/2$ when the number of elements is $N$ ($N\geq 4$).
\end{theorem}

This theorem is based on the idea that if there are $x$ total linearly independent pairings, then the required number of observations is $x$. This algorithm is proved by the following explanation. First, by design, the set of $\tilde{C}_{i,j}$ preserves the total compatibility $C_{\rm{sum}}(S)$ obtained from $C_{i,j}$ for all pairings, and the number of independent $\tilde{C}_{i,j}(\neq0)$ is $(N-1)(N-2)/2$. Therefore, the minimum number of observations is at most $(N-1)(N-2)/2$. Second, the values indicated in Eqs. (2), (3), and (4) represent $(N-1)(N-2)/2$ linearly independent observables, such that the minimum number of observations is at least $(N-1)(N-2)/2$. For these reasons, the minimum number of observations required to know the complete set of compatibilities is $(N-1)(N-2)/2$ when the number of elements is $N$ ($N\geq 4$).

\section{Combining Algorithm}
Based on $\tilde{C}_{i,j}$ obtained by the observation algorithm, we can compute the total compatibility $C_{\rm{sum}}(S)$ of all possible pairings $S$. 
However, as discussed in the Introduction, the number of pairings scales up very quickly as a function of $N$.
In this section we re-formulate the pairing problem into a Traveling Salesman Problem (TSP) to realize an efficient combining algorithm. 

\subsection{Traveling Salesman Problem}
TSP concerns finding the route that minimizes the total cost of traveling to a given set of locations, with the cost between each two locations being given. The salesman starts his or her tour from a starting node and visits all other nodes exactly once before returning to the starting node. The complexity of the TSP stems from the large number of possible routes which scales up very quickly with the number of nodes, such that a brute-force solving considering all possible routes is too costly in general.
\subsection{Solving Pairing Problem as a TSP: Pairing-TSP}
\begin{figure}[h]
\centering
\includegraphics[width=8cm]{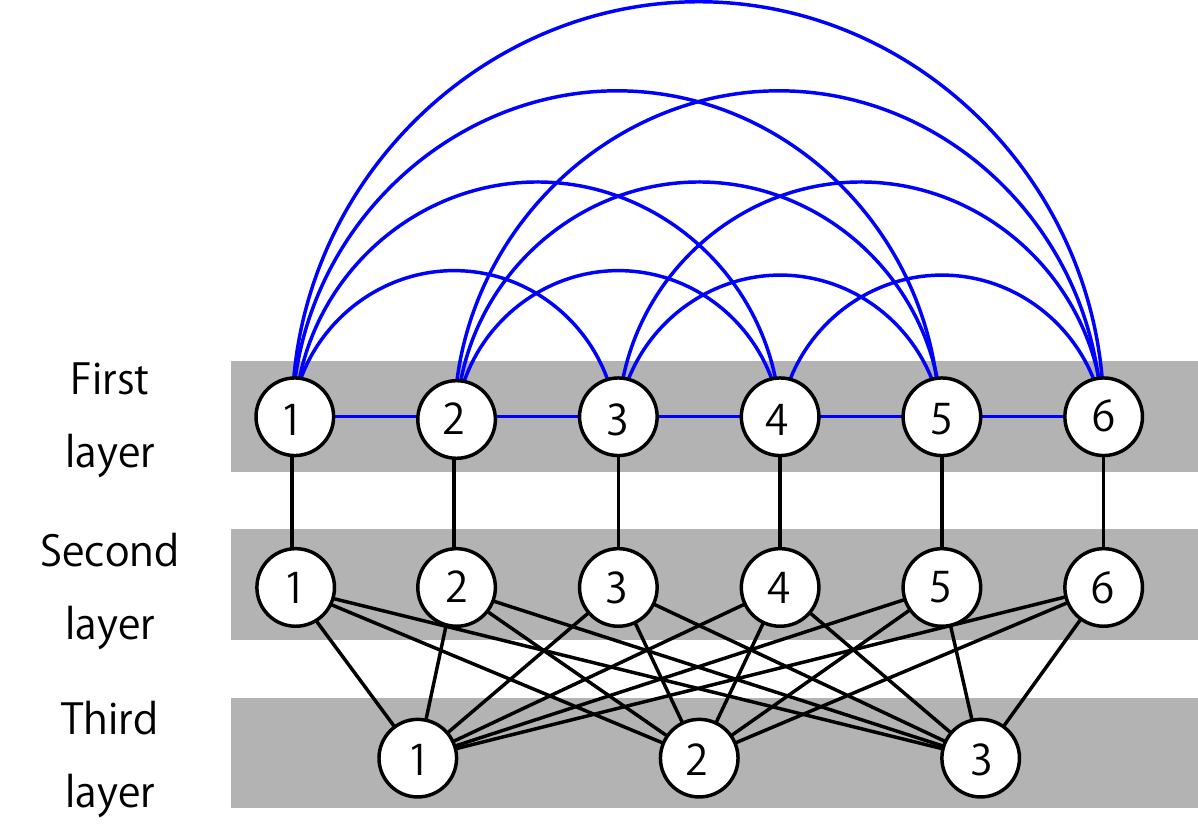}
\caption{The path of the traveling salesman problem in the three-layer graph structure (Pairing-TSP) corresponds to the pairing problem. An example case with the number of elements ($N$) being 6 is illustrated. The first and second layers have $N$ nodes, and the third layer has $N/2$ nodes. All nodes in the first layer are connected with each other. All nodes in the second layer are connected to a different node in the first layer and all nodes in the third layer. By constructing such a three-layer graph structure, the solution to the corresponding TSP problem provides the pairing yielding high compatibility.}
\label{NOMATSP}
\end{figure}
In this study, we transform the problem of heuristically finding the pairing with a large total compatibility into a TSP with a three-layer network structure, which is schematically shown in Fig. \ref{NOMATSP}. 
We call the re-formulated problem \textit{Pairing-TSP}. 

In this Pairing-TSP, we arrange the first and the second layers to have $N$ nodes, while the third layer is configured with $N/2$ nodes.
Let the $N$ nodes of the first and the second layers be indexed with natural numbers ranging from 1 to $N$. 
In the first layer, the cost of the route between the nodes $i$ and $j$ are given by $-C_{i,j}$. 
There is a one-to-one correspondence between the nodes in the first layer and the nodes in the second layer; in other words, there is a unique link between each node $i$ in the first layer and the corresponding node $i$ in the second layer. As the other links between the first and second layers are not permitted, the Pairing-TSP results in a non-complete graph.

Finally, the third layer consists of $N/2$ nodes indexed between $1$ and $N/2$, $N$ being even. Here, the nodes in the second layer and the nodes in the third layer are fully connected. That is, the node $i$ in the second layer is connected with nodes $j$ ($j=1, \cdots, N/2$) in the third layer. Note that the cost of all routes except intra-first-layer links is set to zero. Nevertheless, remember that the salesman must visit all nodes in the second and the third layer too, not just the first layer.
Now we demonstrate that the solution of such Pairing-TSP corresponds to the solution of pairing by noticing the following two inherent constraints. 
\begin{figure}[h]
\centering
\includegraphics[width=8cm]{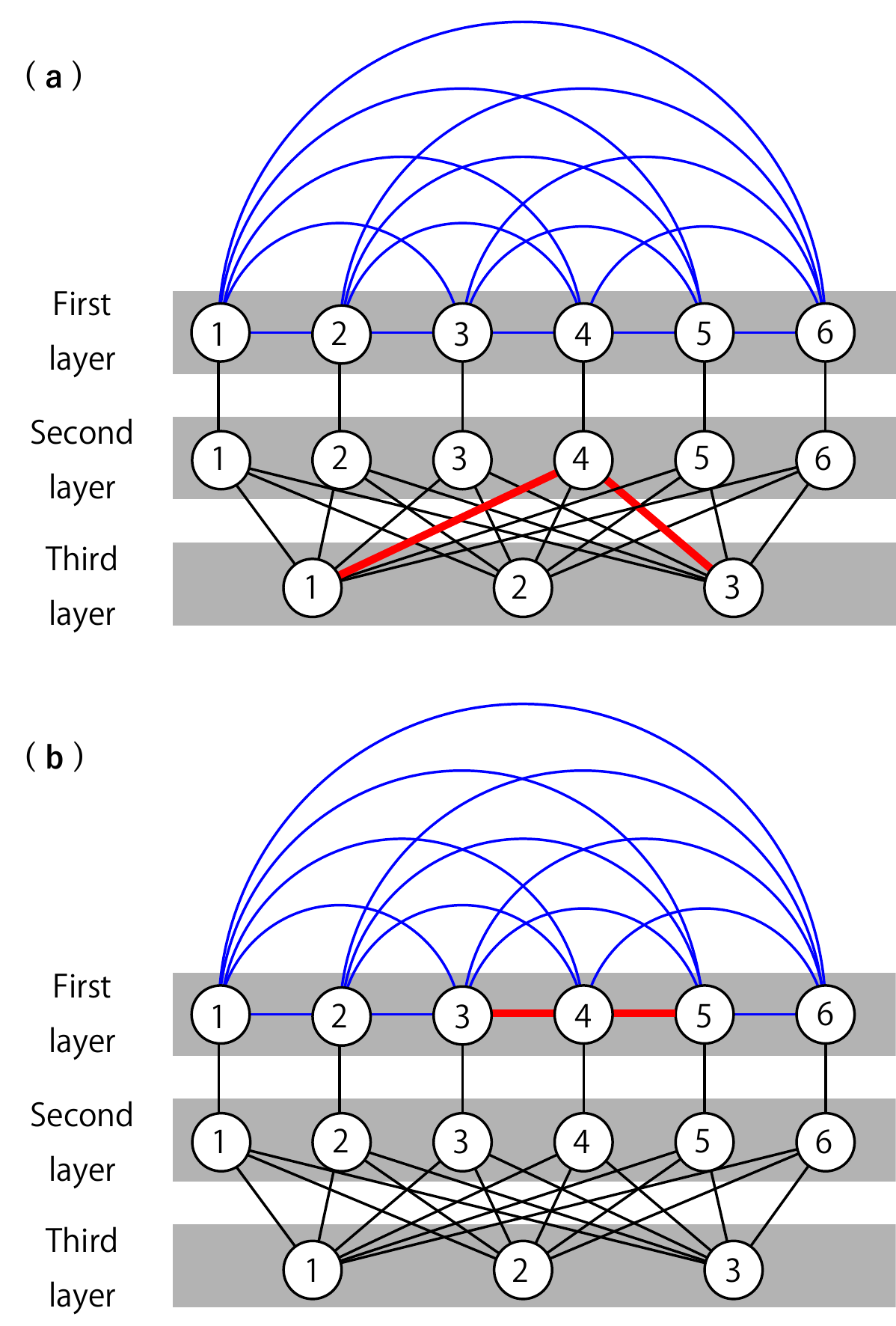}
\caption{A solution to the TSP problem in the three-layer graph structure corresponds to a pairing. This can be explained via forbidden routes illustrated in the following two examples. (a) A route that goes from the third layer to the third layer by passing through the second layer cannot be included in the TSP solution. (b) A route that visits three nodes of the first layer consecutively cannot be included in the TSP solution.}
\label{NOMATSPex_all}
\end{figure}

First, consider a route that goes from a node in the third layer to a node in the second layer and then goes back to the third layer, as shown by the red lines in Fig. \ref{NOMATSPex_all}(a). Such a route fragment cannot be included in the solution of TSP. Each node in the third layer can be connected to at most 2 nodes in the second layer. Therefore, if different nodes in the third layer are connected to the same node in the second layer, there will be at least one node in the second layer that cannot be connected to the third layer. With these reasons, a route fragment such as the red lines in Fig. \ref{NOMATSPex_all}(a) is forbidden.

Secondly, the case of the thick red lines in Fig. \ref{NOMATSPex_all}(b) of three consecutive connections in the first layer cannot be included in the solution of TSP. The reason is if such connections exist, then there has to be the configuration of Fig. \ref{NOMATSPex_all}(a) somewhere. Therefore, by construction, the salesman never visits three consecutive nodes in the first layer; instead, after visiting two nodes in the first layer, the salesman always moves to the second layer. 
Finally, in the solution of Pairing-TSP, the salesman will visit two nodes in the first layer consecutively via visiting the second and the third layer. When the connection between the nodes $i$ and $j$ in the first layer is included in the solution of Pairing-TSP, we consider that elements of $i$ and $j$ are paired. 

Since the summation of the cost along the route of a solution of Pairing-TSP and the total compatibility of the corresponding pairing are opposite in sign, minimizing the cost of Pairing-TSP is equivalent to maximizing the total compatibility $C_{\rm{sum}}(S)$ by appropriate pairing construction. For those reasons, we can guarantee the correspondence between the original pairing problem and Pairing-TSP.

\subsection{Pairing-Nearest Neighbor Method (PNN)}
In solving Pairing-TSP, we propose two algorithms on the basis of existing algorithms for the general TSP. 
\begin{figure}[h]
\begin{algorithm}[H]
    \caption{Pairing-Nearest Neighbor Method (PNN)}
    \label{alg1}
    \begin{algorithmic}[1]
    \REQUIRE Array indexes start at 1
    \STATE \textbf{input}: $C$ ($C$ is the $N \times N$ compatibility matrix, whose element $C[i][j]$ stores compatibility $C_{i,j}$)
    \STATE $q$ ($q$ stores the nodes the salesman visits, and $q[x]$ denotes the node which salesman visits $x$th)
    \STATE $s \leftarrow$ start point in the first layer
    \STATE $q[1] \leftarrow s$
    \STATE $q[\frac{5}{2}N] \leftarrow$ adjacent node of $s$ in the second layer
    \STATE $t \leftarrow 1$
    \WHILE{$t \leq \frac{5}{2}N-2$}
    \IF{$t \mod 5 = 1$}
    \STATE $s \leftarrow$ nearest adjacent node $j\notin q$ of $s$ in the first layer (If there are multiple nearest adjacent nodes, the salesman chooses $s$ with the same probability in them)
    \ELSIF{$t \mod 5 = 2$}
    \STATE $s \leftarrow$ adjacent node of $s$ in the second layer
    \ELSIF{$t \mod 5 = 3$}
    \STATE $s \leftarrow$ adjacent node $j\notin q$ of $s$ in the third layer chosen with the same probabilities
    \ELSIF{$t \mod 5 = 4$}
    \STATE $s \leftarrow$ adjacent node $j\notin q$ of $s$ in the second layer chosen with the same probabilities
    \ELSIF{$t \mod 5 = 0$}
    \STATE $s \leftarrow$ adjacent node of $s$ in the first layer
    \ENDIF
    \STATE $t \leftarrow t+1$
    \STATE $q[t] \leftarrow s$
    \ENDWHILE
    \RETURN $q$
    \end{algorithmic}
\end{algorithm}
\end{figure}
The first one is what we call the pairing-nearest neighbor method, which is referred to as PNN in short hereafter. PNN is a modification of the nearest neighbor method, which is an algorithm to visit the nearest unvisited node from the current node~\cite{NN}. As discussed in Sect. IV.B., a solution of Pairing-TSP does not allow three or more consecutive node visits in the first layer, the salesman needs to go to the second, the third, and the second layer before coming back to the first layer again. If there are multiple least-cost routes to the next node, they are assumed to be chosen randomly with equal probability. This algorithm can obtain an estimated solution with a computational complexity of $\mathcal{O}(N^2)$.
A pseudo-code of PNN is summarized in Algorithm 1. Herein, $q$ denotes the route of the salesman, and $t$ represents the time step of the salesman. Note that there are in total $5N/2$ vertices through which the salesman travels. Lines 4 and 5 specify the start and the end node, respectively. The time step $t$ suggests where the salesman is in the three-layer structure as well as his/her directions to the downward or upward of the layers. There are five kinds of possible movement of the salesman; (1) Move from the 1st layer to the 1st layer. (2) Move from the 1st layer to the 2nd layer. (3) Move from the 2nd layer to the 3rd layer. (4) Move from the 3rd layer to the 2nd layer. (5) Move from the 2nd layer to the 1st layer. In all cases, the destination is chosen only from the unvisited nodes. In this manner, duplicate visits to any node are avoided.

\subsection{Pairing 2-opt Method (P2-opt)}
The second algorithm is what we call the pairing 2-opt method referred to as P2-opt hereafter, which is a modification of the 2-opt method \cite{2opt} to update the initial solution derived by PNN. 
\begin{figure}[h]
\centering
\includegraphics[width=7.5cm]{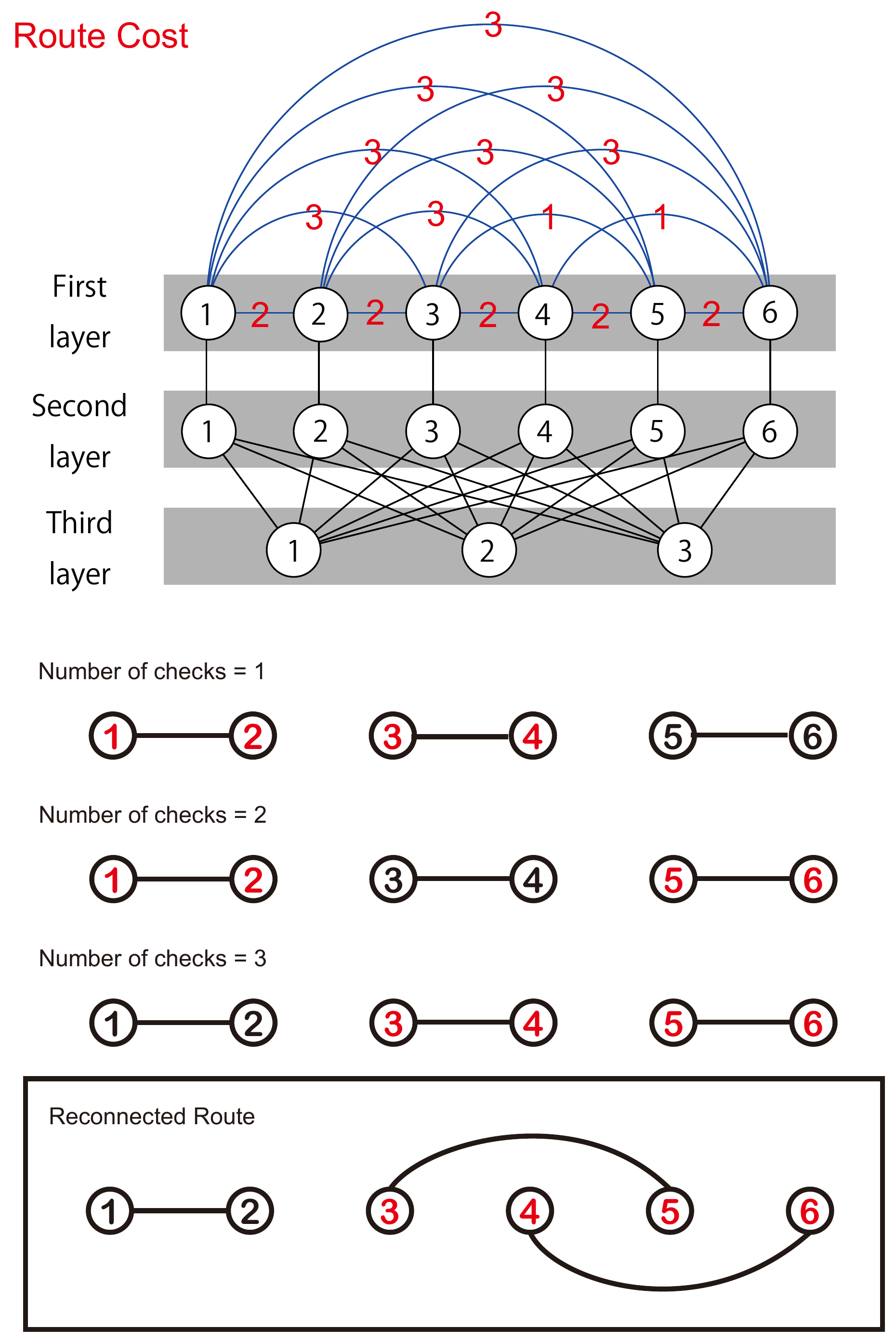}
\caption{Example of P2-opt reconnection when $N=6$. The red numbers overlaid in the connections indicate the cost of connections. At each check, two pairs (or two connections) are considered while all other pairings (or connections) remain the same. Here we consider the pairing combination of $\{\{1,2\}\{3,4\}\{5,6\}\}$. We first examine reconnections concerning $\{1,2\}$ and $\{3,4\}$. In this situation, we check the three alternative pairings or connections $\{\{1,2\},\{3,4\}\}$, $\{\{1,3\},\{2,4\}\}$, and $\{\{1,4\},\{2,3\}\}$. Since $\{\{1,2\},\{3,4\}\}$ is the smallest cost route, the reconnections are not applied. Second, we examine reconnections concerning $\{1,2\}$ and $\{5,6\}$. In this case, we check three alternatives $\{\{1,2\},\{5,6\}\}$, $\{\{1,5\},\{2,6\}\}$, and $\{\{1,6\},\{2,5\}\}$. The reconnection is not adapted again because $\{\{1,2\},\{5,6\}\}$ is the smallest cost route. Third, reconnections about $\{3,4\}$ and $\{5,6\}$ are investigated. The three alternatives are $\{\{3,4\},\{5,6\}\}$, $\{\{3,5\},\{4,6\}\}$, and $\{\{3,6\},\{4,5\}\}$. Here the reconnection to $\{\{3,5\},\{4,6\}\}$ is applied since it yields the minimum cost route. In this case, the number of checks (NOC) is three.}
\label{p2opt_image}
\end{figure}
\begin{figure}[h]
\begin{algorithm}[H]
    \caption{Pairing 2-opt Method (P2-opt)}
    \label{alg2}
    \begin{algorithmic}[1]
    \REQUIRE Array indexes start at 1
    \STATE \textbf{input}: $S$ ($S$ stores which nodes are paired; $S[2k-1]$ and $S[2k]$ are paired for each positive integer $k \leq N$)
    \STATE \textbf{input}: $C$ ($C$ is the $N \times N$ compatibility matrix, whose element $C[i][j]$ stores compatibility $C_{i,j}$)
    \STATE \textbf{input}: $l$ (exchange limit)
    \STATE $t \leftarrow 0$
    \WHILE{$t \leq l $}
    \FOR{$i=1$ to $\frac{N}{2}-1$}
    \FOR{$j=i+1$ to $\frac{N}{2}$}
    \STATE $a\leftarrow C[S[2i-1]][S[2i]]+C[S[2j-1]][S[2j]]$
    \STATE $b\leftarrow C[S[2i-1]][S[2j]]+C[S[2j-1]][S[2i]]$
    \STATE $c\leftarrow C[S[2i-1]][S[2j-1]]+C[S[2j]][S[2i]]$
    \IF{$b=\rm{max}(a,b,c)$}
    \STATE $x\leftarrow S[2i]$
    \STATE $S[2i]\leftarrow S[2j]$
    \STATE $S[2j]\leftarrow x$
    \STATE $t \leftarrow t+1$
    \STATE break all for-loops
    \ELSIF{$c=\rm{max}(a,b,c)$}
    \STATE $x\leftarrow S[2i]$
    \STATE $S[2i]\leftarrow S[2j-1]$
    \STATE $S[2j-1]\leftarrow x$
    \STATE $t \leftarrow t+1$
    \STATE break all for-loops
    \ENDIF
    \IF{$i=\frac{N}{2}-1$ and $j=\frac{N}{2}$}
    \RETURN $S$
    \ENDIF
    \ENDFOR
    \ENDFOR
    \ENDWHILE
    \RETURN $S$
    \end{algorithmic}
\end{algorithm}
\end{figure}
The original 2-opt method compares the original and one alternative route and updates the current solution by reconnecting some of the nodes so that the total cost decreases \cite{2opt}. Conversely, in Pairing-TSP, there are three ($\{\{i,j\},\{k,l\}\},\{\{i,k\},\{j,l\}\},\{\{i,l\},\{k,j\}\}$) possible combinations for 2 given pairs of 4 nodes. Therefore, the proposed P2-opt compares the costs of three routes. If the compatibility is not improved by recombining any of the pairs, the algorithm terminates. Fig. \ref{p2opt_image} illustrates the reconnection procedure of the proposed P2-opt with an example of pairing when $N = 6$. A pseudo-code of P2-opt is shown in Algorithm 2. The three alternatives are represented by lines 8 to 10. In P2-opt, the rewiring is considered only on the first layer among these three alternatives. This rewiring never introduces duplicate visits. Note that the connections involving the 2nd and 3rd layers have zero cost for the salesman. Therefore, any rewired route in the first layer, which is a pairing $S$, provides a certain route for the salesman in the three-layer graph structure.

\section{Simulation}
\subsection{Problem Setting}
We constructed the compatibility set $C_{i,j}$ by generating uniform random numbers between 0 and 10000. A total of 100 different $C_{i,j}$ sets were generated for each setting, and the average over different settings was examined. Note that each set of compatibilities $C_{i,j}$ is reconstructed here following the observation algorithm based on the construction of $\tilde{C}_{i,j}$ described in Sect. III. We want to compare the performance of PNN versus random pairing, evaluate how much P2-opt can improve a solution found by PNN through additional rewiring steps, and how the performance gain depends on the number of rewirings as introduced in Sect. IV.

\subsection{Performance Indicator for the Derived Pairing}
Let $C_{\rm{sum}}(S)$ be the total compatibility that corresponds to the pairing $S$ derived through the combining algorithm. The larger $C_{\rm{sum}}(S)$ and the closer it is to the global maximum, the better it is. 
To quantify the performance of the combining algorithm in terms of how far $C_{\rm{sum}}(S)$ is from the maximum, we define $P$ as a performance indicator with the following formula:
\begin{equation}
P(S)=\frac{C_{\rm{sum}}(S)-\frac{N}{2}C_{\rm{min}}}{\frac{N}{2}C_{\rm{max}}-\frac{N}{2}C_{\rm{min}}}
\label{eq:definition_p}
\end{equation}
where $N$ is the number of nodes in the first layer, $C_{\rm{max}}$ is the upper limit value of $C_{i,j}$, and $C_{\rm{min}}$ is the lower limit value of $C_{i,j}$. In this simulation, $C_{\rm{max}}=10000$ and $C_{\rm{min}}=0$. $P$ ranges from 0 to 1 and represents the relative distance of the current pairing from the theoretical minimum or maximum possible values for $C_{\rm{sum}}(S)$, 0 being for the absolute worst and 1 for the absolute best pairing, respectively.

\subsection{Performance of PNN and P2-opt}
\begin{figure}[h]
\centering
\includegraphics[width=8cm]{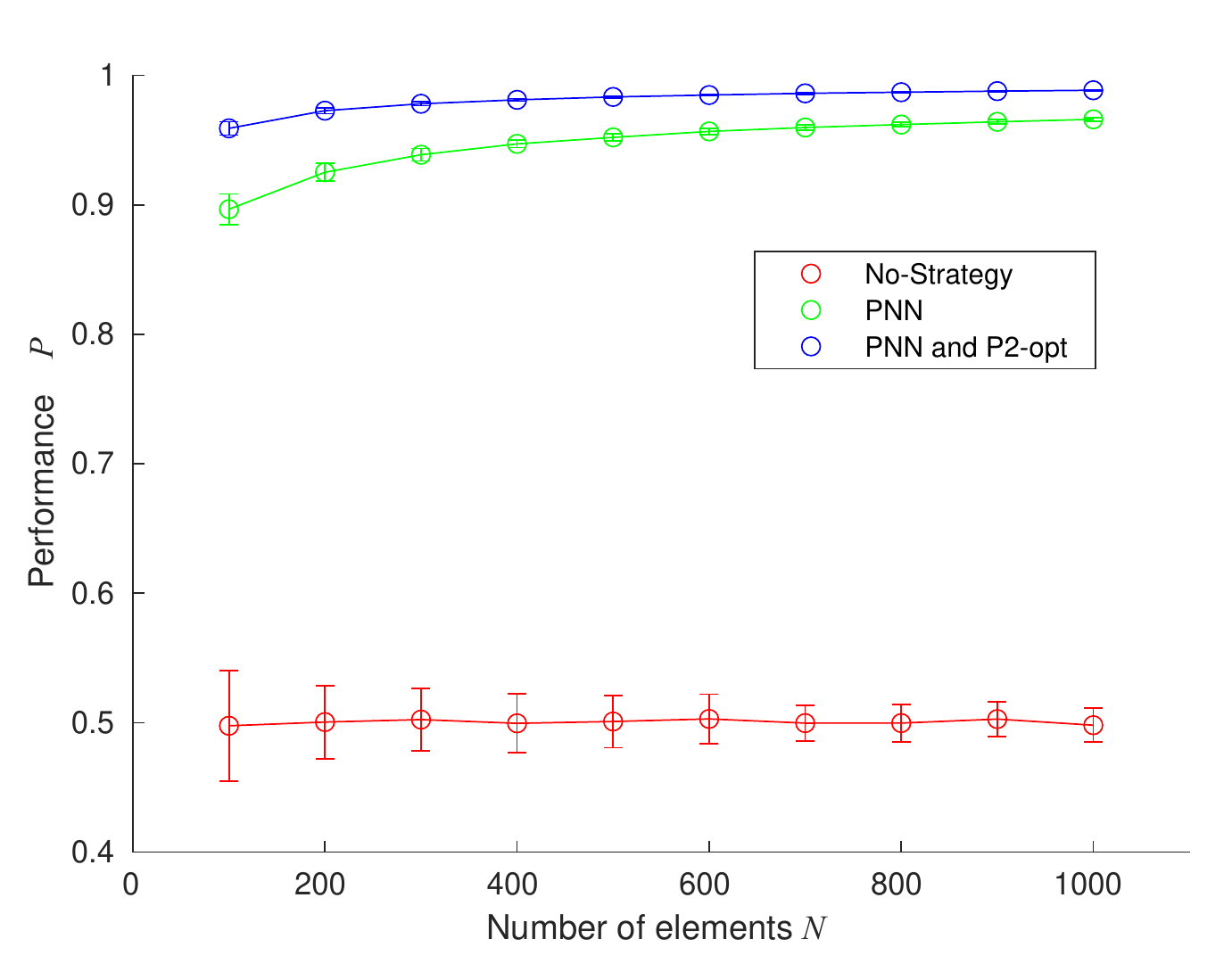}
\caption{Performance comparison by the index $P$ among (a) No-Strategy, (b) PNN, and (c) PNN and P2-opt methods as a function of the number of elements $N$. We can observe that PNN greatly improves the performance, and P2-opt provides additional enhancements.}
\label{Performance}
\end{figure}

We conducted a performance comparison between (a) No-Strategy, (b) PNN, (c) PNN and P2-opt as a function of the number of elements $N$ from 100 to 1000, as summarized in Fig. \ref{Performance}. Herein the exchange limit $l$ was fixed to be 600. "No-Strategy" indicates random selection of the route in the first layer. "PNN and P2-opt" means that we get an initial solution by PNN and update solution by P2-opt. 

The performance of No-Strategy is roughly 0.5 regardless of $N$, as expected by the definition of $P$ in Eq. \eqref{eq:definition_p}. The pairing of the proposed strategies, PNN and P2-opt, reaches a performance index greater than 0.9. Furthermore, we can confirm that P2-opt processing enhanced the solution of PNN. Furthermore, the standard deviation tends to be smaller for (c) PNN and P2-opt, (b) PNN, and (a) No-strategy, in that order. Regarding the relationship between the number of elements $N$ and performance, the performance of both (b) PNN and (c) PNN and P2-opt improves as the number of elements increases. The standard deviation tends to decrease for all three methods as the number of elements increases.

\subsection{Effect of P2-opt}
\begin{figure}[h]
\centering
\includegraphics[width=8.4cm]{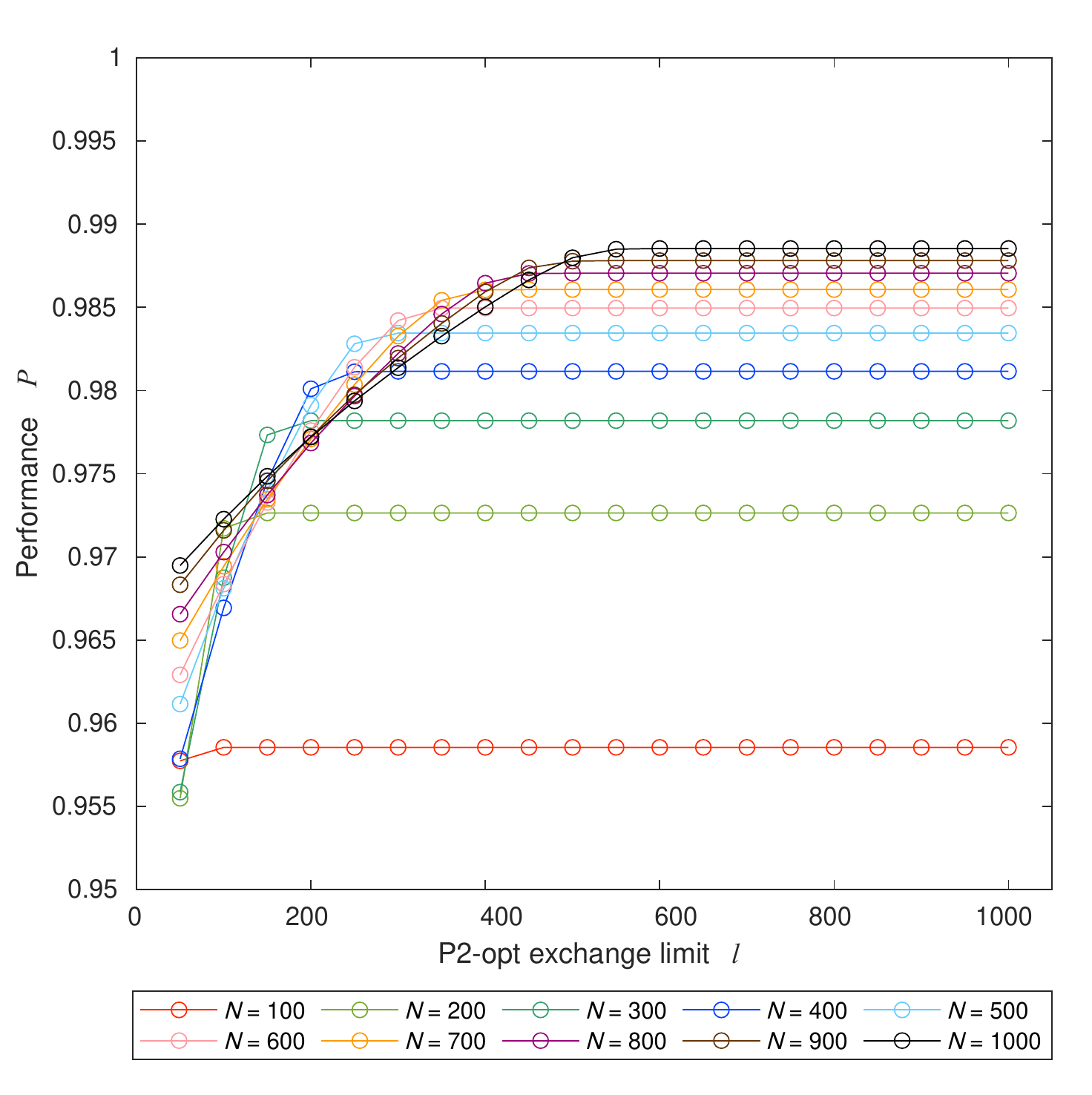}
\caption{Performance evaluation of the proposed P2-opt algorithm, i.e. rewiring of connections in the first layer of the Pairing-TSP. The performance ($P$) increases as a function of exchange limit ($l$) in the P2-opt algorithm. The colors indicate the different numbers of elements ranging from 100 to 1000.}
\label{p_parameters}
\end{figure}

As described in Sect. IV.D., P2-opt aims at reducing the total cost of a TSP route by locally exchanging connections. To examine the effect of such an exchange, here we set an upper limit to the number of exchanges in P2-opt, which we define by the P2-opt exchange limit denoted by $l$. 
Fig. \ref{p_parameters} shows the evolution of $P$ as a function of $l$ for different element numbers $N$ from 100 to 1000 in intervals of 100, each point representing the average among 100 different compatibility sets. This result highlights two trends: first, $P$ saturates beyond a certain limit $l$; second, as $N$ increases, increasing $l$ improves the performance until a new saturation level. Indeed, when $N=100$, the performance reached its maximum value with $l=100$, whereas $P$ monotonically increases until $l=600$ when $N=1000$. These observations demonstrate that a sufficient exchange limit exists depending on the number of first-layer nodes of the given problem.
\subsection{Number of checks of P2-opt}
In the P2-opt algorithm, two pairs of the current pairing are compared at every turn, and the nodes are reconnected accordingly if the rewiring improves the total compatibility (Fig. \ref{p2opt_image}). Here, the order in which the pairs are checked is round-robin, meaning that each time the pairs are reconnected, they are rechecked from the beginning. Therefore, there is a possibility of double-checking, meaning that certain reconnections are re-calculated.  That is to say, there is a room for further accelerating the algorithm in reducing the number of checks. 

\begin{figure}[h]
\centering
\includegraphics[width=8.4cm]{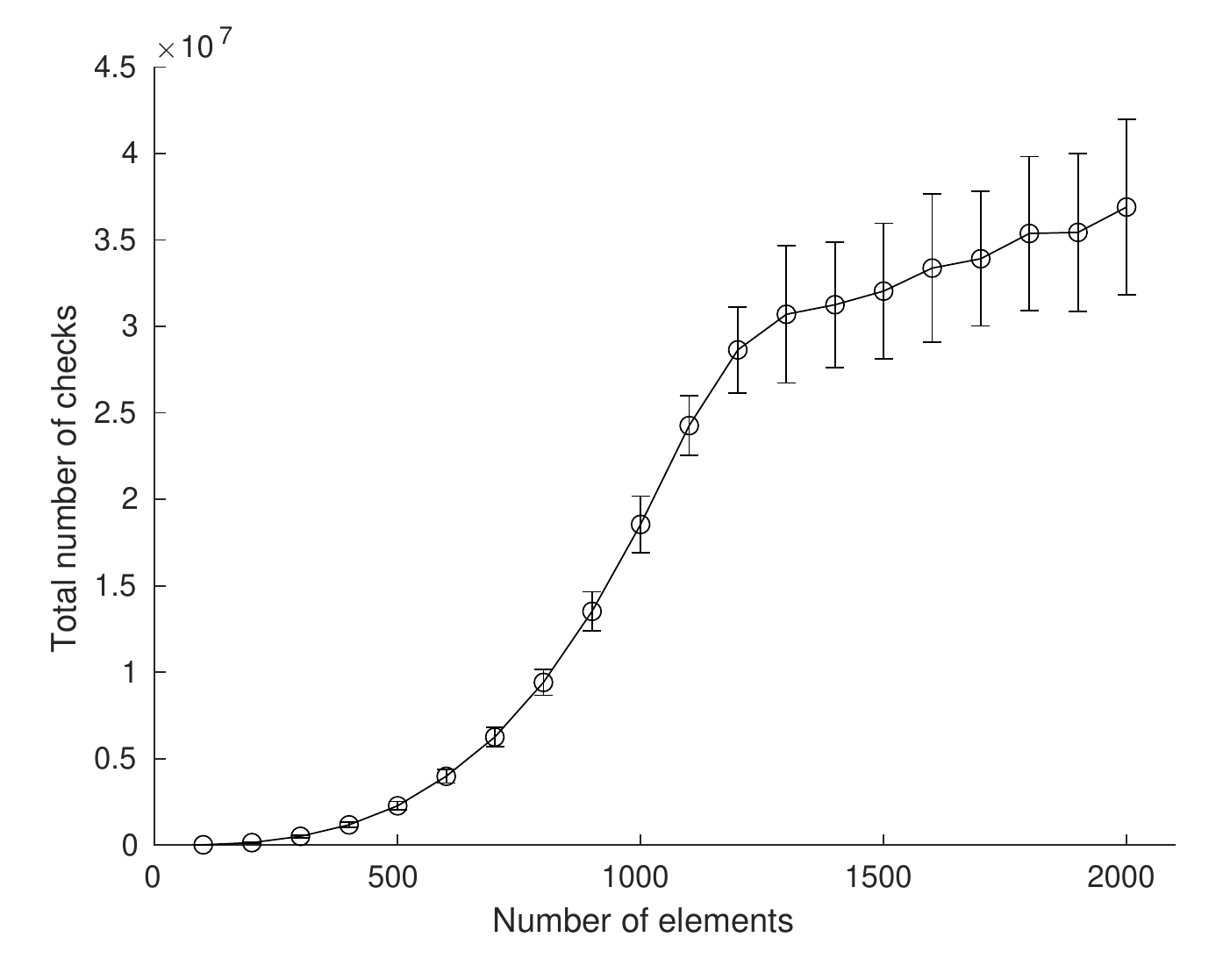}
\caption{Evaluation of the computational cost in P2-opt. A total number of checks (NOC), which quantifies how often compatible pairs are compared, is evaluated as a function of the number of elements ($N$). 100 different compatibility sets were examined for each $N$. The graph shows the average and standard deviation.}
\label{count}
\end{figure}

In the meantime, the computation cost of the P2-opt algorithm represents how often compatible pairs are compared, which we call the \textit{number of checks} (NOC). The circular marks and their associated error bars in Fig. \ref{count} represent the mean and the standard deviation of the NOC, respectively, when the number of elements $N$ ranges from 100 to 2000 in intervals of 100. For each $N$, 100 different compatibility sets were examined. The exchange limit $l$ was given by 600 regardless of $N$. However, when P2-opt achieves the local maximum pairing and the algorithm terminates, then the total number of exchanges is actually less than $l$. \\
\begin{figure}[h]
\centering
\includegraphics[width=8cm]{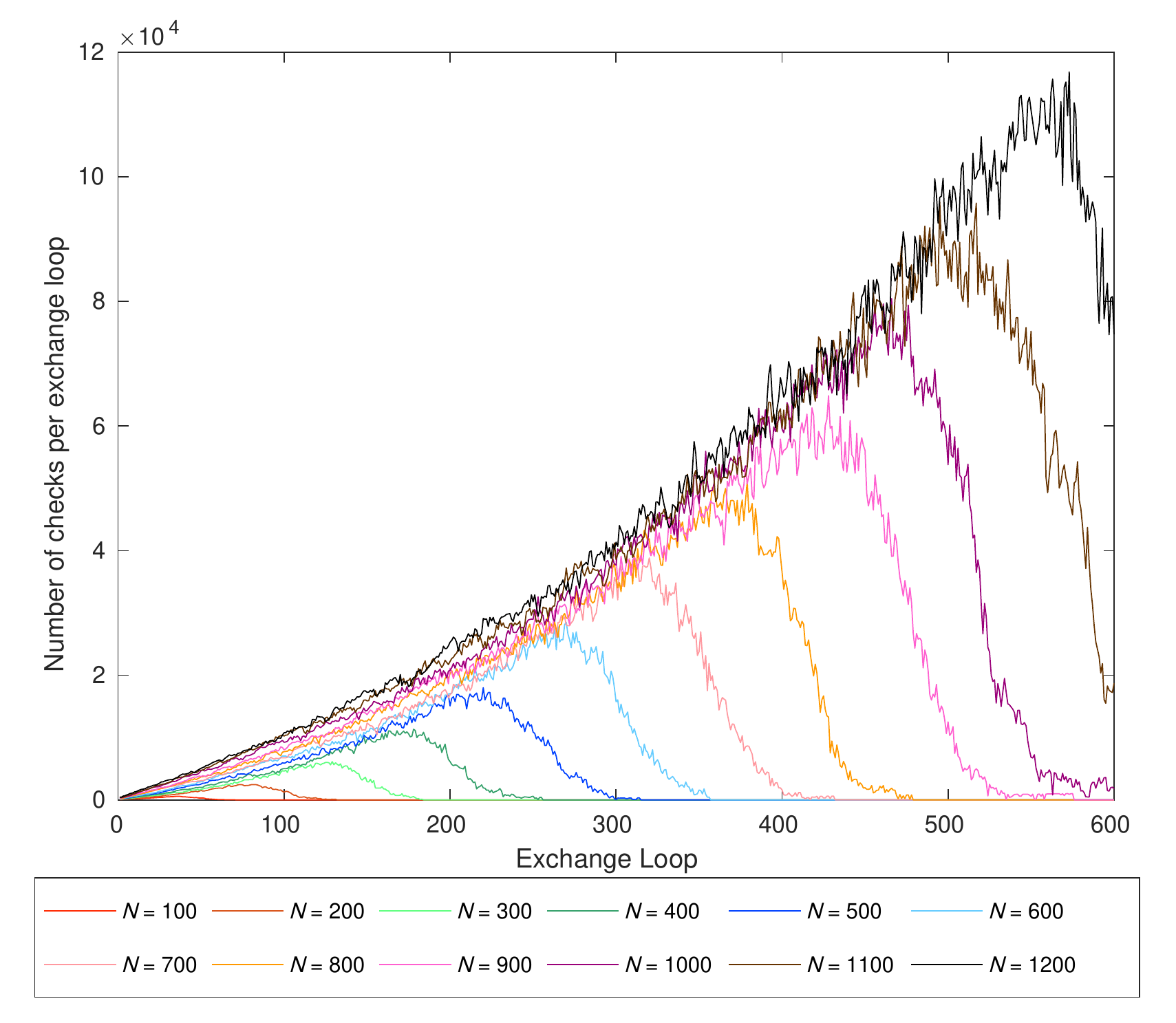}
\caption{Analysis of the underlying mechanism of P2-opt. The number of actually conducted checks per exchange loop is evaluated as the progress of the algorithm. Here, the exchange limit ($l$) is set as 600. For each $N$, the graph shows the average value over 100 different compatibility sets.}
\label{p_parameters_count}
\end{figure}
\indent From Fig. \ref{count}, we can observe several trends. Clearly, the NOC increases as the number of elements increases. However, the slope flattens when the number of elements is greater than approximately 1200. Furthermore, the standard deviation gets larger when the number of elements goes beyond 1200.\\
\indent To examine the inherent mechanisms behind such tendencies, we analyzed the time evolution of the NOC per exchange loop. The curves in Fig. \ref{p_parameters_count} represent the evolution over exchange loops of the NOC regarding compatibility settings whose number of elements ranges from 100 to 1200 in intervals of 100, averaged over 100 different compatibility sets for each setting. The P2-opt exchange limit $l$ was fixed at 600. From Fig. \ref{p_parameters_count}, we observe that the average NOC initially increases as number of exchange loops elapses. Initially, any rewiring may improve the total compatibility; hence the NOC per exchange loop is small. As the number of exchanges increases, rewiring may not necessarily improve the total compatibility because the calculated route may already be in a good solution. Therefore, the NOC until actual rewiring happens increases. Beyond a certain point, the calculated route has a relatively low cost; therefore, the NOC grows until P2-opt has converged, but becomes 0 once P2-opt has converged. In Fig. \ref{p_parameters_count}, 100 trials were simulated for each $N$ and averaged over, such that the NOC gradually decreased after some point because the number of converged trials steadily increased.

Indeed, in the case of $N = 500$, the NOC becomes almost zero when exchange loop is 300. 
Similarly, in the case of $N=1000$, the NOC becomes very small when the total number of exchanges is 600. In the case of $N=1200$, however, the NOC is large, approximately $8 \times 10^{4}$ when the total number of exchanges is 600. That is to say, the search for a better solution may be insufficient. Such an observation is consistent with the change of the slope in Fig. \ref{count} induced at $N = 1200$. In other words, when $N$ is small, the variance is small because a sufficiently low-cost route solution has been obtained, whereas when $N$ is greater than 1000, the $l$ is insufficient, and so the variance becomes large, and the slope of the graph against $N$ is slow.

\subsection{Comparison of computational costs}
In this section, we discuss the computational complexity of each method. First, the total number of possible pairings is $(N-1)!!$. Therefore, the computational complexity by enumeration is $(N-1)!!$, and the number of observations required is also $(N-1)!!$. On the other hand, the number of observations needed for the proposed observation algorithm is $\mathcal{O}(N^2)$, and the computational complexity of the proposed combining algorithm is $\mathcal{O}(N^2)$ for PNN and at most $\mathcal{O}(lN^2)$ for P2-opt.

\section{Conclusion}
In this study, we propose an algorithm for efficiently and heuristically determining a pairing that provides large total compatibility among entities, which lies in a process at the heart of some of the latest information and communications technologies such as non-orthogonal multiple access (NOMA) in wireless networks, matching problems in economics, among others. We identify two main phases to optimize the pairing: observation and combination. One of the main hypotheses of this study is that one can only observe the total compatibility for any given one pairing. In the meantime, the number of all possible pairing pairings grows as $(N-1)!!$, where $N$ is the number of entities. Therefore, efficient strategies to measure the compatibility among elements are essential. We demonstrate that the minimum number of observations to know the complete set of all compatibilities is smaller than the total number of combinations of this set. This finding does not depend on the combining phase. Also, by exploiting the exchange relationships inherent in the problem, we propose an efficient algorithm scaling as $\mathcal{O}(N^2)$ to observe all compatibilities among elements. 
In the combining phase, we demonstrate that the derivation of the best pairing is equivalent to solving a traveling salesman problem (TSP) in a three-layer graph structure, which we name Pairing-TSP. We propose two heuristic approaches to efficiently resolve Pairing-TSP: the pairing-nearest neighbor (PNN) and the pairing 2-opt (P2-opt) methods, both of which exploit unique characters inherent in the architecture of Pairing-TSP. Numerical simulations confirm the principles of the algorithms. 
In summing up, the present study first proposed an algorithm to estimate the compatibility among elements only via the total compatibility with minimal observations. Then, through the insight that the pairing problem is equivalent to solving a special class of TSPs, we demonstrate heuristic methods to accomplish pairing efficiently. We consider that the contents herein contribute to achieving more efficient pairing than conventional methods, especially for the case of a large number of users in NOMA systems, as well as other pairing applications.
We expect our findings to be applicable also to social systems such as social networking services and education.

\section{Effect of Initial Node in PNN}
In the PNN, traveling starts from a node in the first layer. Here we examined the effect of the starting node on the resultant pairing performance. More specifically, we analyzed the standard deviation of the performance indicator $P$ defined in Sect. V.B. while changing the starting node through all $N$ nodes in the first layer. In the simulations, $N$ was given from 100 to 1000 with a 100 interval, while 100 types of compatibilities were prepared for each  given $N$. We calculated standard deviations for $N$ initial nodes for each of the 100 compatibility sets. Then, we averaged all 100 standard deviations for each $N$.
\begin{figure}[h]
\centering
\includegraphics[width=8.4cm]{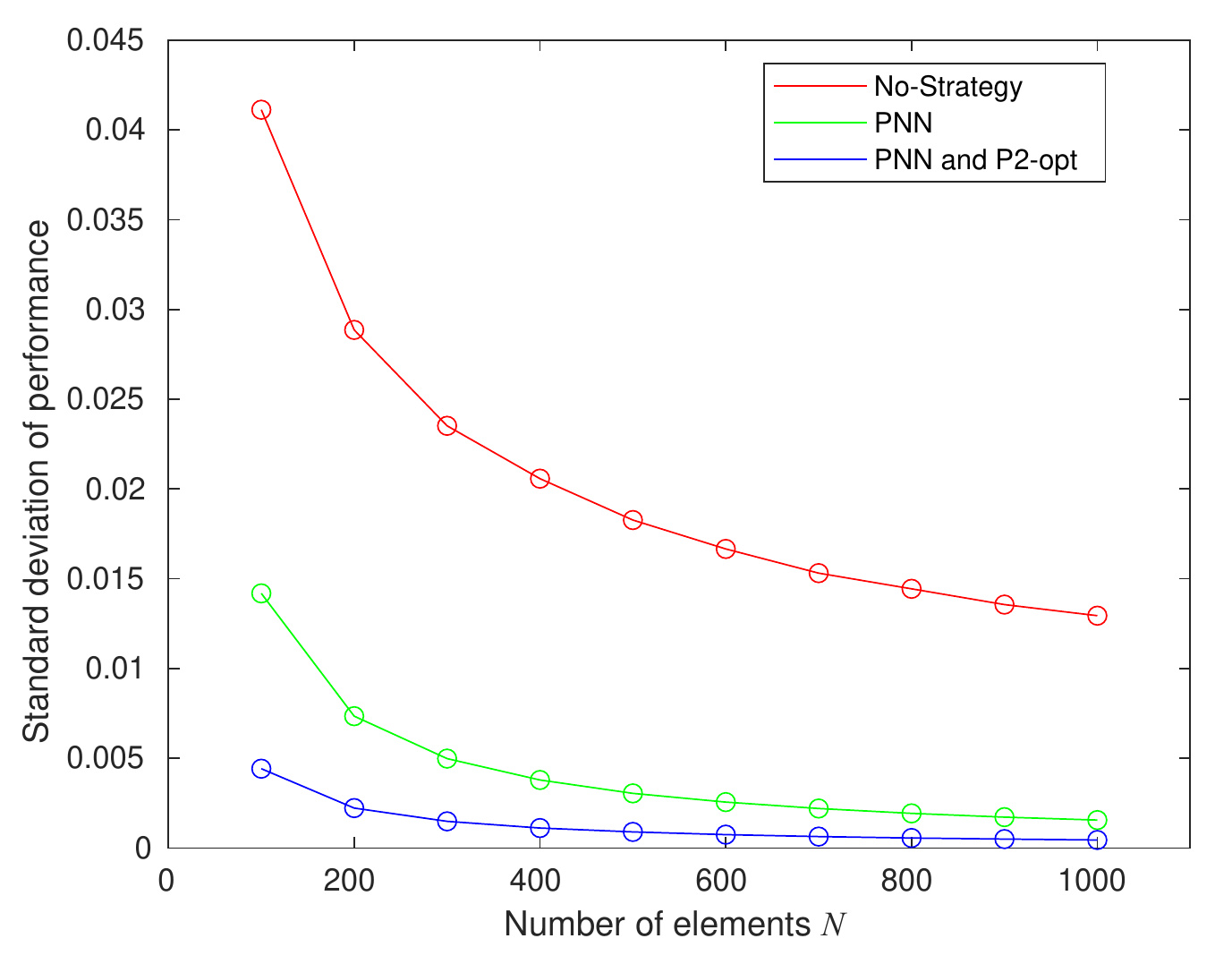}
\caption{The standard deviation of the performance for each method as a function of $N$ when the initial point is changed. $N$ is set from 100 to 1000 and we prepare 100 types of compatibilities for each $N$.}
\label{initial_dependency}
\end{figure}
The red, green, and blue circular marks in Fig. \ref{initial_dependency} show the standard deviation of the performance indicator $P$ as a function of the number of elements $N$ when the pairing attribution was conducted with completely random strategy (or No-Strategy), PNN, and PNN and P2-opt, respectively. We can observe that the standard deviation decreases as the number of elements increases for all methods. In particular, the dependence of the performance on the initial node of PNN and P2-opt is smaller than that of No-Strategy and PNN. 
Since the maximum standard deviation is smaller than $0.015$ when $N=100$ in the case of PNN and PNN and P2-opt, we can conclude that the initial node selection in PNN has a negligible effect on the resultant pairing quality.

\newpage
\end{document}